\documentclass[preprint,showpacs,preprintnumbers,amsmath,amssymb,superscriptaddress,floatfix]{revtex4}
\usepackage{graphicx}
\usepackage{color}
\usepackage{dcolumn}% Align table columns on decimal point
\usepackage{bm}% bold math
\begin{document}
\title{Glass transitions in 1, 2, 3, and 4 dimensional binary Lennard-Jones systems }

\author{Ralf Br\"uning}
\email[]{rbruening@mta.ca, Tel.: (506) 364-2587, Fax.: (506) 364-2583}
\author{Denis A. St-Onge}
\affiliation{Physics Department, Mount Allison University, Sackville, New Brunswick,
Canada E4L 1E6}
\author{Steve Patterson}
\affiliation{Physics Department, Mount Allison University, Sackville, New Brunswick,
Canada E4L 1E6}
\affiliation{present address: Department of Physics and Atmospheric Science, Dalhousie University, Halifax, NS, Canada B3H 3J5}
\author{Walter Kob}
\affiliation{Laboratoire des Collo\"ides, Verres et Nanomat\'eriaux, UMR5587, 
Universit\'e Montpellier II and CNRS, 34095 Montpellier Cedex, France}


\date{\today}

\begin{abstract}
We investigate the calorimetric liquid-glass transition by performing
simulations of a binary Lennard-Jones mixture in one through four
dimensions. Starting at a high temperature, the systems are cooled to
$T=0$ and heated back to the ergodic liquid state at constant rates.
Glass transitions are observed in two, three and four dimensions as a
hysteresis between the cooling and heating curves.  This hysteresis
appears in the energy and pressure diagrams, and the scanning-rate
dependence of the area and height of the hysteresis can be described
by power laws.  The one dimensional system does not experience a glass
transition but its specific heat curve resembles the shape of the
$D\geq 2$ results in the supercooled liquid regime above the
glass transition. As $D$ increases, the radial distribution functions
reflect reduced geometric constraints. Nearest-neighbor distances become
smaller with increasing $D$ due to interactions between nearest and
next-nearest neighbors.  Simulation data for the glasses are compared
with crystal and melting data obtained with a Lennard-Jones system
with only one type of particle and we find that with increasing $D$
crystallization becomes increasingly more difficult.

\end{abstract}

\maketitle

\section{\label{Introduction} Introduction}

As a glass-forming liquid is cooled, its relaxation time increases
very rapidly and at sufficiently low temperatures, the relaxation
time eventually exceeds the time scale of cooling. Thus,
provided that crystallization is avoided, the system forms
a glass at a temperature $T_g$ and due to the kinetic effects
the non-equilibrium state of this glass depends on its thermal history
\cite{ritland54,yang87,limbach88,struik78,johari89,miyagawa89,bruning92,hodge94,levelut95,vollmayr96,liu01,buchholz02,lee06,leroux07,streit08}.
In recent years much theoretical effort has been made to gain a better understanding
of the mechanism(s) responsible for the dramatic slowing down of the
relaxation dynamics as well as to investigate the aging dynamics of the
system once the system has fallen out of equilibrium~\cite{zarzycki91,gotze92,ediger96,debenedetti97,gotze99,cugliandolo03,das04,binder05}. Most of
these studies have been done for three dimensional systems since this
correspond to the overwhelming majority of real experiments, with some
notable exceptions in which the (experimental) glass transition has been investigated
in quasi-two dimensional systems~\cite{marcus99,konig05}. Fewer investigations
have been devoted to the question to what extend the phenomenon of
the glass transition depends on the dimensionality of the system since
experimentally it is rather difficult to change the dimensions without
modifying the interactions. Nevertheless, such a study is of interest,
since it allows one to estimate the role of the local geometry on the
glass-forming ability of the system. Although it is evident that with
increasing dimensionality $D$ the geometric constraints decrease, it is
difficult to estimate this tendency within an analytical calculation in
a more quantitative way~\cite{thalmann02,parisi06,tarzia07}. One possibility to
address the problem is to use computer simulations to determine, with a
given interaction potential, the dependence of the thermodynamic
and structural properties of the system as a function of $D$.

In the present paper we study therefore how the glass
transition of a binary-mixture Lennard-Jones (BMLJ) system
depends on its dimensionality $D$.  In the past it has been
shown that such a system shows, for $D=3$, many properties of real
glass formers and thus it can serve as a good model for glass-forming
systems~\cite{kob94,kob95a,kob95b,vollmayr96,nauroth97,gleim98,sciortino99,berthier04,murarka04,flenner05,berthier07}.
Here we will focus on the details of the glass transition as characterized by the
specific heat measurements, we will examine the structure of this system in its glassy state
and we will compare the results with experimental data. Note that
here we will not discuss the relaxation dynamics and we refer to
Refs.~\cite{ranganathan94,hurley95,perera99,doliwa00,santen00,bayer07} in which
such investigations have been done for two-dimensional systems.

Section II of this paper discusses the details of the simulation,
including the type of system and the method by which it is
studied. Section III describes the results of the simulations, starting
with microscopic properties of the system, moving to the macroscopic
properties. Concluding remarks are given in Section IV.

\section{\label{model} Model and Details of the Simulations}

\begin{figure}[top]	           
%goose/denis/ljcnh/pgc
\includegraphics[scale=1.0]{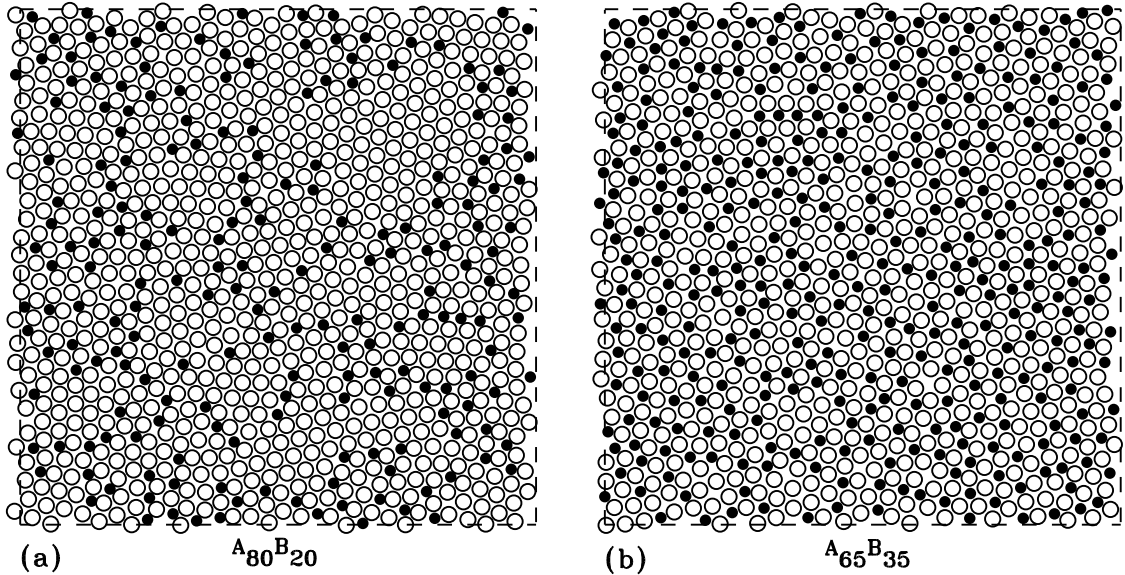}

\caption{\label{pGlass}
Typical configurations of the A$_{80}$B$_{20}$ (a) and A$_{65}$B$_{35}$
(b) systems for $D=2$ at $T=0$. These systems have been cooled to $T=0$
with cooling rates $\gamma = -1.0 \times 10^{-5}$ and $\gamma=-1.0 \times
10^{-7}$, (a) and (b) respectively. Open and filled disks represent
A and B particles, respectively. Dashed lines indicate the (virtual)
boundary of the simulation box.
}
\end{figure}
	
Following previous work, we consider a binary mixture of particles
A and B, all having the same mass, $m$ \cite{kob94,kob95a,kob95b}.
We extend the usual three-dimensional BMLJ molecular dynamics simulation
to one ($D=1$) through four ($D=4$) spatial dimensions. A number of
particles, $N$, is placed inside a box with edge length $L$ and constant
volume $V=L^D$.  As in previous work, periodic boundary conditions are
imposed~\cite{frenkel96}. The interactions between particles are given by
the Lennard-Jones potential, $U_{\alpha\beta}(r)~=~4\epsilon_{\alpha\beta}
[(\sigma_{\alpha\beta}/r)^{12}-(\sigma_{\alpha\beta}/r)^{6}]$, where
$\alpha,\beta \in \{$A, B\,$\}$,  $\sigma_{\textrm{AA}}=1.0$,
$\epsilon_{\textrm{AA}}=1.0$, $\sigma_{\textrm{AB}}=0.8$,
$\epsilon_{\textrm{AB}}=1.5$, $\sigma_{\textrm{BB}}=0.88$
and $\epsilon_{\textrm{BB}}=0.5$ ~\cite{kob94}. Here $r$
is defined through $r^2= \sum_{k=1}^Dx_k^2$, where  $x_k$ is
the $k^\textrm{th}$ cartesian component of the inter-particle
separation.  Following common practice, the potential is truncated
and shifted at $r =2.5\sigma_{\alpha\beta}$~\cite{kob94,frenkel96}.
Reduced units are used, with $\sigma_{\textrm{AA}}$ being the unit
of length,  $\epsilon_{\textrm{AA}}$ the unit of energy, $\left(
m\sigma_{\textrm{AA}}^2/48\epsilon_{\textrm{AA}} \right)^{1/2}$ the
unit of time ~\cite{kob94}, and Boltzmann's constant, $k_B$, is set
equal to one.  The temperature, $T$, is controlled by a Nos\'e-Hoover
thermostat~\cite{frenkel96} with an effective mass of 48 reduced
units.	For the molecular dynamics of the particles, the equations of
motion are integrated using the Verlet algorithm with a time step of
0.02~\cite{kob94}. The pressure, $P$, is monitored using the virial
theorem, $PV=NT+D^{-1}\sum_{i<j}{\textbf{f}_{ij}\cdot\textbf{r}_{ij}}
$, where $\textbf{f}_{ij}$ is the force and $\textbf{r}_{ij}$  the
separation between particles $i$ and $j$.

Part of the simulations were carried out for the composition
A$_{80}$B$_{20}$ in three dimensions, a system that has been studied
extensively~\cite{kob94,kob95a,kob95b,nauroth97,gleim98,sciortino99,berthier04,murarka04,flenner05,berthier07}.
For $D=1,2,3$, and $4$, the composition A$_{65}$B$_{35}$ was selected
because it is, unlike A$_{80}$B$_{20}$, stable against crystallization
for $D=2$ at the cooling rates employed here.  All simulations begin with
the system in equilibrium at a sufficiently high initial temperature.
The temperature is lowered at the rate	$-\gamma$ to $T=0$ and then
increased back to the initial temperature at the rate $\gamma$, where
$ \gamma=1.0 \times 10^{-3}$, $1.0 \times 10^{-4}$, and $1.0 \times
10^{-5}$.  Further simulations with $\gamma=1.0\times 10^{-6}$ have
been performed for $D=2,3$, and for $D=2$ additional simulations were
carried out at $\gamma=1.0\times 10^{-7}$.  The A$_{80}$B$_{20}$ system
was studied with $\gamma$ in the range from $\gamma=1.0\times 10^{-3}$ to
$\gamma=1.0\times 10^{-7}$.  For $D=1,2$, and $3$ the number of particles
is $N=1000$, while $N$ is 2000 for $D=4$.  To increase the statistical
significance of the results, data are averaged over independent runs
with different initial configurations.	At the starting temperatures the
relaxation times are very short, equilibrium is rapidly attained and the
statistically independent starting configurations are readily obtained.
The starting configurations for $D=1$ simulations are random sequences
of A and B particles.  In general the results represent averages over
100 runs.  Due to computation time constraints, only 20 runs are used
for the $D=4$, A$_{65}$B$_{35}$ system as well as for A$_{80}$B$_{20}$
with $\gamma \le 1.0 \times 10^{-6}$.

The particle density $\rho$ for the A$_{80}$B$_{20}$
simulations is the same as in previous work with $L=9.4$ in $D=3$,
i.e. $\rho=N/L^D=1.204$~\cite{kob94}.  To establish a common reference
point for all A$_{65}$B$_{35}$ systems at different dimensionalities,
the system volume was chosen such that the simulation pressure is
approximately zero when the temperature reaches $T=0$ upon cooling
at $\gamma=-1.0 \times 10^{-4}$.  The resulting box edge lengths are
$L=1002.5$, $29.34$, $8.88$ ($N=1000$) and $5.68$ ($N=2000$) for $D=1,2,3$
and $4$, respectively.  { It has been shown that the bulk properties
of $D=3$ BMLJ systems emerge with as few as 65 particles
\cite{doliwa03}.}

Further simulations with only A particles (A$_{100}$) were carried out
in order to identify the features that distinguish the vitreous from the
crystalline state.  Box sizes, adjusted such that $P \approx 0$ for the
crystalline state at $T=0$, are 1119 ($N=1000$), 32.7 ($N=1000$), 9.85
($N=1000$), and 6.0639 ($N=2048$) for $D=1$ to 4.  Temperature scanning
rates ranged between $\gamma=\pm1.0 \times 10^{-3}$ and $\pm1.0 \times
10^{-4}$.  The $D=2$ and $D=3$ systems crystallize spontaneously upon
cooling whereas the $D=4$ remained in a metastable state. Therefore
the $D=4$ system was heated from an initially prepared fcc crystal
consisting of $4^4$ unit cells that each contain 8 particles with 24
equidistant neighbors~\cite{pfender04}.

\section{\label{theoretical} Results and Discussion}
	
Since we are interested in the phenomenon of the glass transition, we have
to consider the stability of the model system against crystallization.
Figure~\ref{pGlass}a shows a typical configuration of the particles for
the A$_{80}$B$_{20}$ composition, in two dimensions, cooled to $T=0$
at a rate of $-1.0 \times 10^{-5}$. As observed previously for a similar
system~\cite{ballone02}, the configuration contains areas of hexagonally
crystallized A particles in a matrix of amorphous AB material, i.e. the
system which is a good glass-former in three dimensions is crystallizing
in two dimensions.  In order to suppress these hexagonal A crystals for
$D=2$, we selected the composition A$_{65}$B$_{35}$. Figure~\ref{pGlass}b
shows a typical configuration of a A$_{65}$B$_{35}$ system which has
been cooled to $T=0$ at a rate of $-1.0 \times 10^{-7}$, the slowest rate
employed in this study.  The structure appears to be fully amorphous, as
required. In the following we will first discuss the results obtained for
the A$_{65}$B$_{35}$ composition.  Commonalities and differences between
A$_{65}$B$_{35}$ and A$_{80}$B$_{20}$ for $D=3$ will be considered at
the end of this section.

\begin{figure}[top]	           
\includegraphics[scale=0.7]{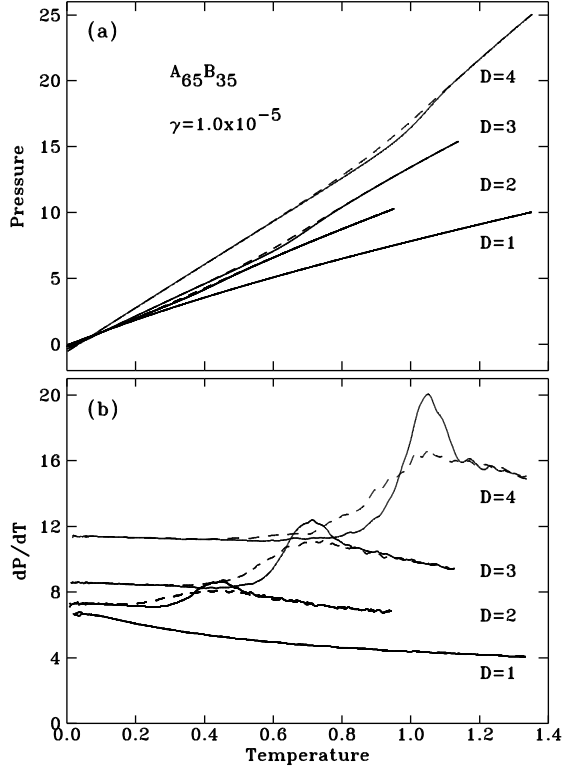}
\caption{\label{pressure}
(a) Temperature dependence of the pressure of the A$_{65}$B$_{35}$ systems for $D=1$
to $D=4$ upon cooling (dashed lines) followed by heating (solid
lines) at $\gamma= \mp 1.0 \times 10^{-5}$.  { (b) Temperature derivatives
of the same pressure data.}
}
\end{figure}

In Fig.~\ref{pressure}(a) we show the temperature dependence of the pressure
for different values of $D$. The $P(T)$ curves are approximately linear,
and their slope increases with the dimensionality. Recall that the particle
density is adjusted for each dimension such that $P\approx0$ as the
temperature reaches $T=0$  upon cooling at $\gamma =-1.0\times 10^{-4}$
(see Sec.~\ref{model}).  Actual pressures obtained after cooling to
$T=0$ at this rate are between $-0.10$ and $0.00$.  Cooling the system
more slowly results in a more relaxed glass state, i.e. a more efficient
packing of the particles and thus a lower pressure at $T=0$.  For the rate
$\gamma=-1.0\times10^{-5}$  (Fig.~\ref{pressure}), pressures range from
$-0.59$ ($D=4$) to $0.00$ ($D=1$).  Pressures at $T=0$ are small compared
with pressures at $T_g$, with $|P(T=0)|$ less that $5\%$ of $P(T_g)$,
as required for consistency.  

The figure also shows that there is a hysteresis between the
cooling curve and the heating curve for $D \ge 2$, which becomes more
pronounced as $D$ increases.  {This hysteresis is seen more
clearly in the temperature derivative of the pressure, $dP/dT$ (Fig.~\ref{pressure}(b)).}
As it will be discussed below, the glass
transition temperatures  for the data shown in Fig.~\ref{pressure} are
$T_g=0.33,~0.58$, and $0.89$ for $D=2,3,4$, respectively. These values
for $T_g$ were obtained by first calculating the fictive temperature of
the system, $T_f(T)$, as defined by Tool~\cite{tool46}.  The procedure is
based on analytic approximations for the specific heat of the supercooled
liquid \cite{sciortino00} and the glass (further details are given below).
In the liquid state the fictive temperature equals the temperature,
while in the glass state it becomes frozen at a finite value.  We set
$T_g = T_f(0)$, the limiting value of the fictive temperature as the
glass stops evolving upon cooling to low temperatures.

Above the temperature range of this hysteresis loop, the system is
in equilibrium and the pressures upon cooling and heating coincide.
A hysteresis is a hallmark of physical glasses in the glass transition
range~\cite{bruning96} and the glass transition found here is
qualitatively consistent with previous observations of a glass
transition in cooling curves obtained with a $D=3$, A$_{80}$B$_{20}$
BMLJ system~\cite{vollmayr96}. Last not least we remark that the $D=1$
system does not show any sign of a glass transition since in one
dimension this Lennard-Jones model is not sufficiently frustrated to
form a glass. (However, other one-dimensional models can show a glass
transition, see, e.g., Ref.~\cite{kob90}.)

\begin{figure}[top]
%location goose/denis/ljcnh/prdfaa_pap
\includegraphics[scale=0.9]{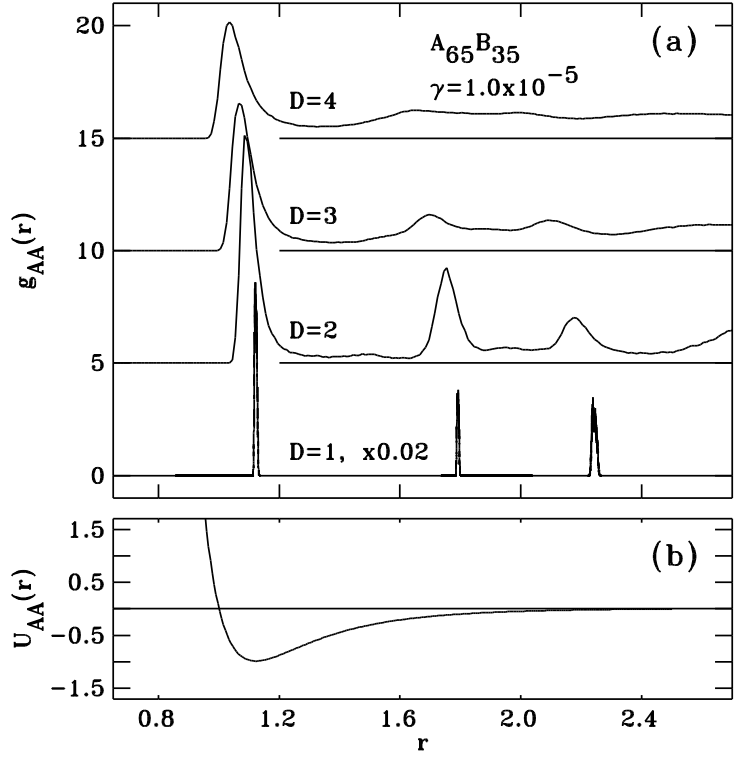}
\caption{\label{prdfaa}
(a) Radial distribution of A--A pairs for A$_{65}$B$_{35}$
cooled to $T=0$ at the rate $-1.0 \times 10^{-5}$. Data for $D=1$ are
multiplied by 0.02.  Data for $D \ge 2$ are shifted up successively by 5.
(b)  Lennard-Jones potential for A--A interaction, truncated and shifted
at $r=2.5$.
}
\end{figure}	

\begin{figure}[top]
%location goose/denis/ljcnh/prdfab_pap
\includegraphics[scale=0.9]{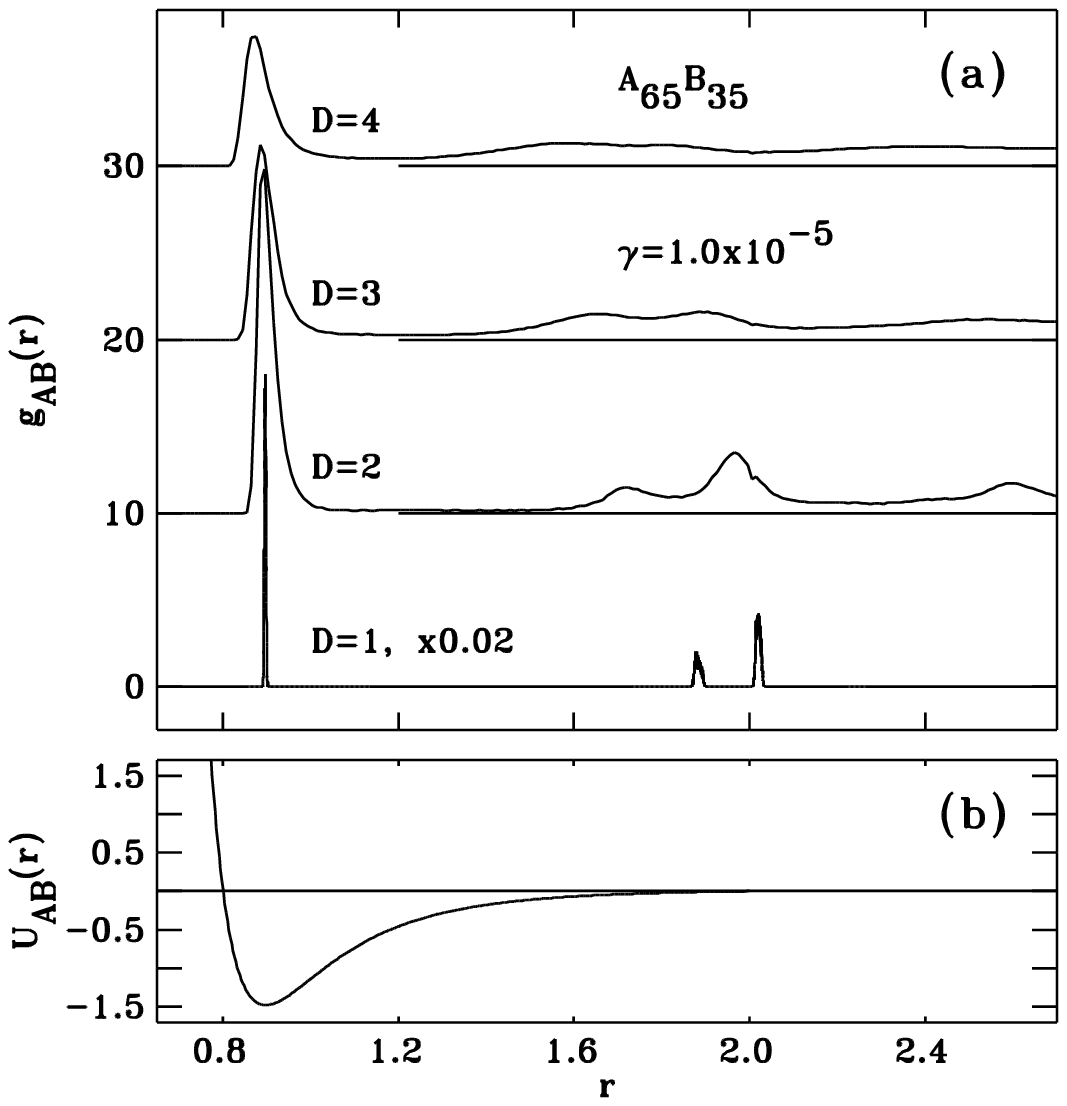}
\caption{\label{prdfab}
(a) Radial distribution of A--B pairs for A$_{65}$B$_{35}$
cooled to $T=0$ at the rate $-1.0\times10^{-5}$.  Data for $D=1$ are
multiplied by 0.02.  Data for $D \ge 2$ are successively shifted up by
10. (b) Lennard-Jones potential for A--B interactions, truncated and
shifted at $r=2.0$.
}
\end{figure}

\begin{figure}[top]
%location goose/denis/ljcnh/prdfbb_pap
\includegraphics[scale=0.9]{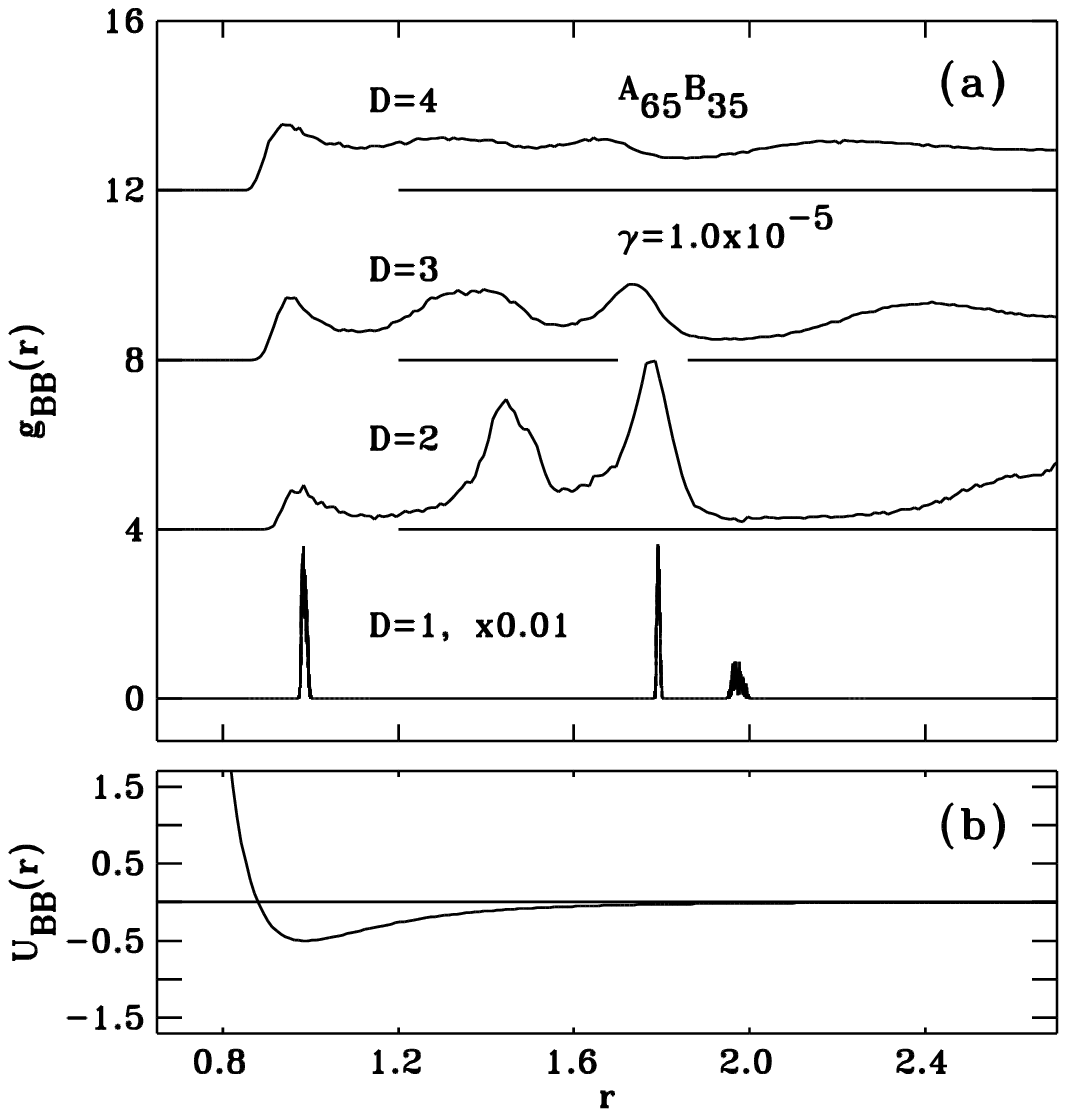}
\caption{\label{prdfbb}
(a) Radial distribution of B--B pairs for A$_{65}$B$_{35}$
cooled to $T=0$ at the rate $-1.0\times10^{-5}$.  Data for $D=1$ are
multiplied by 0.01.  Data for $D \ge 2$ are successively shifted up
by 4. (b) Lennard-Jones potential for B--B interactions, truncated and
shifted at $r=2.2$.
}
\end{figure}	

In order to characterize the local structure of the particles it is useful
to consider the radial pair distribution functions, $g_{\alpha\beta}(r)$,
defined by~\cite{hansen06}

\begin{equation}
g_{\alpha \beta}(r) = \frac{L^D}{F(r,D) N_{\alpha}N_{\beta}} \sum_{i \in \{\alpha\}}
\sum_{j \in \{\beta\}} \langle  \delta (r-r_{ij})\rangle \quad ,
\label{eq_gdr}
\end{equation}

\noindent
where $N_\alpha$ is the number of particles of type $\alpha$, $r_{ij}$
is the distance between particle $i$ and $j$, and the factor
$F(r,D)$ normalizes $g_{\alpha\beta}(r)$ to 1.0 for large $r$,
i.e. $F(r,1)=2$, $F(r,2)=2\pi r$, $F(r,3)=4\pi r^2$, and $F(r,4)=2 \pi^2 r^3$.

Figures~\ref{prdfaa}a, \ref{prdfab}a, and \ref{prdfbb}a show the A--A,
A--B and B--B distributions, respectively, at $T=0$. For $D=1$ one finds
a first peak in $g_{\alpha\beta}$ at a distance $\sigma_{\alpha\beta}
2^{1/6}$, i.e.  at the location of the minimum in the corresponding
pair potential (shown in panel b of the figures). This is due to
the fact that we have adjusted the pressure to be zero at $T=0$. The
second and third A--A peaks in Fig.~\ref{prdfaa}a for $D=1$ are at
$r=1.80$ and $r=2.25$ and correspond to \underline{A}B\underline{A} and
\underline{A}A\underline{A} elements of the particle chain, respectively,
where the underlined letters indicate the atom types considered in the
pair distribution. For $g_{\rm AB}(r)$ for $D=1$, chain elements of
the types \underline{A}B\underline{B} and \underline{A}A\underline{B}
give rise to the peaks at $r=1.89$ and $r=2.03$, respectively.
The $D=1$ B--B radial distribution peaks are due to, in order of
increasing $r$, \underline{BB}, \underline{B}A\underline{B} and
\underline{B}B\underline{B} chain elements.

For $D=2$, the peak at $r=1.76$ corresponds to A--A next-nearest
neighbors that are separated by a nearest neighbor particle of type B,
while the peak at $r=2.20$ correspond to A particles which are separated
by a nearest neighbor particle of type A. The $D=2$ peaks at $r=1.73$
and $r=1.96$ in the A--B distribution, Fig.~\ref{prdfab}, can be
interpreted in a similar fashion and correspond to correlations with
two A and two B nearest neighbors, respectively.  In $g_{\rm BB}(r)$
a new B--B peak arises from the long edge of rectangles formed by four
B particles surrounding an A particle, see Fig.~\ref{pGlass}(b).

We also mention that a notch can be seen in $g_{\textrm{AB}}$ for
$D=2$ at $r=2.0$, the point where the force is discontinuous due to
the truncation of the potential~\cite{frenkel96}.  It is unlikely that this
discontinuity has a significant impact on the results.

The functions $g_{\alpha\beta}(r)$ also show a well
defined nearest neighbor peak for $D=3$ and $4$, but the peaks
at larger distances for these dimensions are much
less pronounced than the ones found for $D=1$ and 2, and it becomes
difficult to associate them with particular arrangements of particles. Note
that also the first nearest neighbor peak becomes broader with increasing
$D$ since the typical distance between neighboring particles will,
at large $D$, be strongly influenced by the type and the number of
the particles that are their common nearest neighbors. For the A--A
correlation the increase of $D$ will, e.g., make it possible that two
A particles share an increasing number of B particles as first nearest
neighbors and, since the A--B interaction is strongly attractive, thus
decrease the nearest neighbor distance of such an A--A pair. Alternatively
an A--A nearest neighbor pair that shares many A particles as first
nearest neighbors, will have a distance that is somewhat larger than
the average nearest neighbor A--A distance.

This mechanism for reducing the geometric constraints between
particles that are nearest neighbors affects of course also the second,
third,... -nearest neighbor configurations. This is the reason why with
increasing $D$ the radial distribution functions become less
structured at a given $r$. In particular the location of the second,
third,...-nearest neighbor peaks will shift to smaller distances and also
the minima between consecutive peaks will be be less pronounced. This
change in the geometry has the effect that the pressure of the system
increases since the second nearest neighbor particles move to distances
in which the potential is steeper/more attractive and hence the virial
increases.

\begin{figure}[top]	           
% puffin:ralf/everything/ljcnh/pstack_kiss
\includegraphics[scale=0.9]{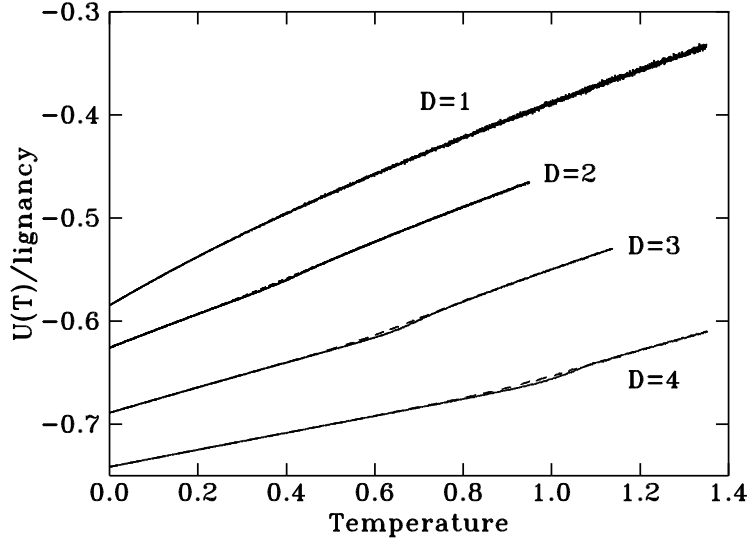}
\caption{\label{pstack_E}
Potential energy per particle versus temperature for A$_{65}$B$_{35}$ upon
cooling (dashed lines) followed by heating (solid lines) at $\gamma=\mp
10^{-5}$.  Potential energies are divided by the number of neighbors in
a closely packed structure (lignancy = 2, 6, 12 and 24 for $D=1$, 2, 3,
4, respectively).
}
\end{figure}

\begin{figure}[top]	           
% puffin:ralf/everything/ljcnh/pcrystu
\includegraphics[scale=0.9]{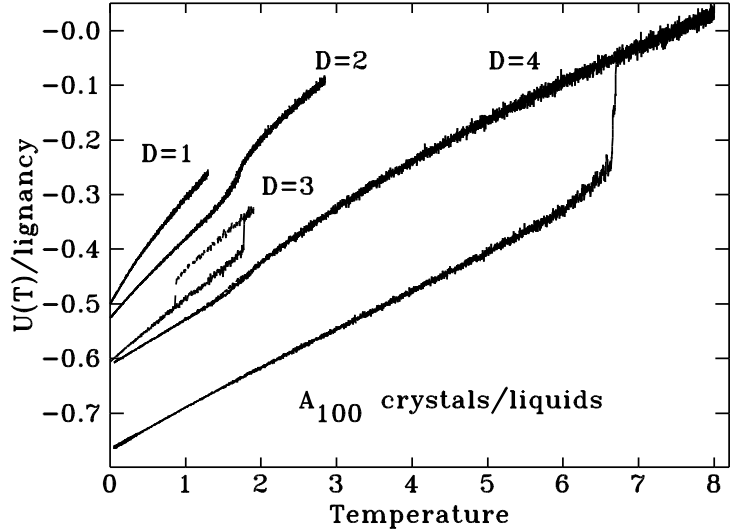}
\caption{\label{pcryst_U}
Potential energy per particle versus temperature for systems containing
only type A particles upon cooling (dashed lines) followed by heating
(solid lines) at $\gamma=\mp 10^{-4}$ for $D = 1$, 2 and 3.  Data for
$D=4$ were obtained at $\gamma=\pm 10^{-3}$ by heating and melting
a closed-packed four-dimensional face-centered cubic crystal (lowest
curve), followed by cooling and heating. Potential energies are divided
by the number of neighbors in a closely packed structure (lignancy = 2,
6, 12 and 24 for $D=1$, 2, 3, 4, respectively).
}
\end{figure}

Having discussed the influence of the dimensionality on the structure we
now present the results regarding the glass transition. One convenient
method to investigate this transition in real glasses are specific heat
measurements. We will consider the temperature
dependence of the specific heat and energy of the BMLJ system. (See
Refs.~\cite{vollmayr96,liu01,buchholz02,lee06,leroux07,vollmayr95,vollmayr96b}
for a discussion how other quantities depend on the cooling rate.)
Figure~\ref{pstack_E} shows the potential energy per particle for
A$_{65}$B$_{35}$ as a function of temperature.  For ease of comparison
the potential energies are scaled by the number of nearest neighbors
in closed-packed sphere structures.  These lignancy values are 2, 6,
12 and 24 for $D=1$ to $D=4$ \cite{pfender04}.  At $T=0$, $P \approx 0$ the interaction
energy of an isolated pair of A particles is 0.5. From Fig.~\ref{pstack_E}
we recognize that for the A$_{65}$B$_{35}$ system the actual values are
lower due to the higher A--B binding energy and the contributions of
the next-nearest neighbor interactions.  Total energies per particle
(not shown) are negative, indicating bound states.  The slopes of the
$U(T)/{\rm lignancy}$ curves decrease with increasing $D$, reflecting
enhanced cohesion and stability of the glass and liquid states
per neighbor-pair.

First-order melting and crystallization are distinct from the glass
transition. This is shown in Fig.~\ref{pcryst_U} for constant-volume
systems composed entirely of A particles. Box sizes were adjusted such
that $P\approx 0$ for the crystalline state at $T=0$, and in order to
compare the curves for the different values of $D$ we have divided them
by the corresponding lignancy. (Note that since the binary system is
a good glass-former, its crystalline structures are very complex and
not really known~\cite{middleton01,fernandez03}.) For $D=1$, the $T-$dependence
of the potential energy of the chain similar to that
of the random A$_{65}$B$_{35}$ chain, and no phase transition
occurs for $T>0$, as expected. In this case $U(T=0)/2$ is 0.5, as
expected, since next-nearest neighbor interactions are insignificant.
The two-dimensional system crystallizes readily upon cooling near $T=1.7$,
and upon reheating melting nearly coincides with the crystallization.
Upon cooling the $D=3$ system stays in a (metastable) supercooled state
down to a temperature $T\approx 0.9$ at which it crystallizes. The melting
of the resulting crystal is observed at around $T=1.8$, i.e. at about
twice the temperature of crystallization. (Note that the exact values
for melting and crystallization are probably affected by finite size
effects. Furthermore one should recall that the resulting ordered
structure does not have a long range positional ordering. However,
these two issues are not that relevant here.)

Interestingly we found that for $D=4$ the pure A system would not
crystallize spontaneously. Therefore we assembled a dense-packed
crystalline face-centered cubic structure with $4^4$ unit cells of eight
particles (2048 particles altogether, each with 24 equidistant nearest
neighbors) \cite{pfender04}. Figure~\ref{pcryst_U}, lowest curve, shows
that this structure melts at $T \approx 6.6$.  Once melted, the $D=4$
system does not crystallize upon cooling but instead forms a glass
near $T=1.5$ (heating and cooling data are shown).  We conclude that
with increasing $D$ spontaneous crystallization becomes increasingly
inhibited, i.e. the glass-forming ability increases with $D$. This trend
is in agreement with the conclusions we drew in the discussion of the
radial distribution functions (Figs.~\ref{prdfaa}, \ref{prdfab}, and
\ref{prdfbb}), i.e. that the number of possible local packings increases
rapidly with $D$ and this entropic factor will make crystallization
more difficult.

From Figure~\ref{pcryst_U} we also recognize that at $T=0$ the potential energy per
lignancy decreases monotonically with $D$ for liquids as well as for the
glasses, in agreement with the trend observed for the binary mixture,
Fig.~\ref{pstack_E}. This results reflects the fact that the next-nearest
neighbor atoms move closer to the central atom as $D$ increases and
hence lower their potential energy. The liquid state is characterized
by a strong negative curvature of the $U(T)$ curves, while the curvature is
much less for the crystal and glass states for $D=2$, 3 and 4.

\begin{figure}[top] 
% goose/denis/ljcnh/pstack_r4b
\includegraphics[scale=0.9]{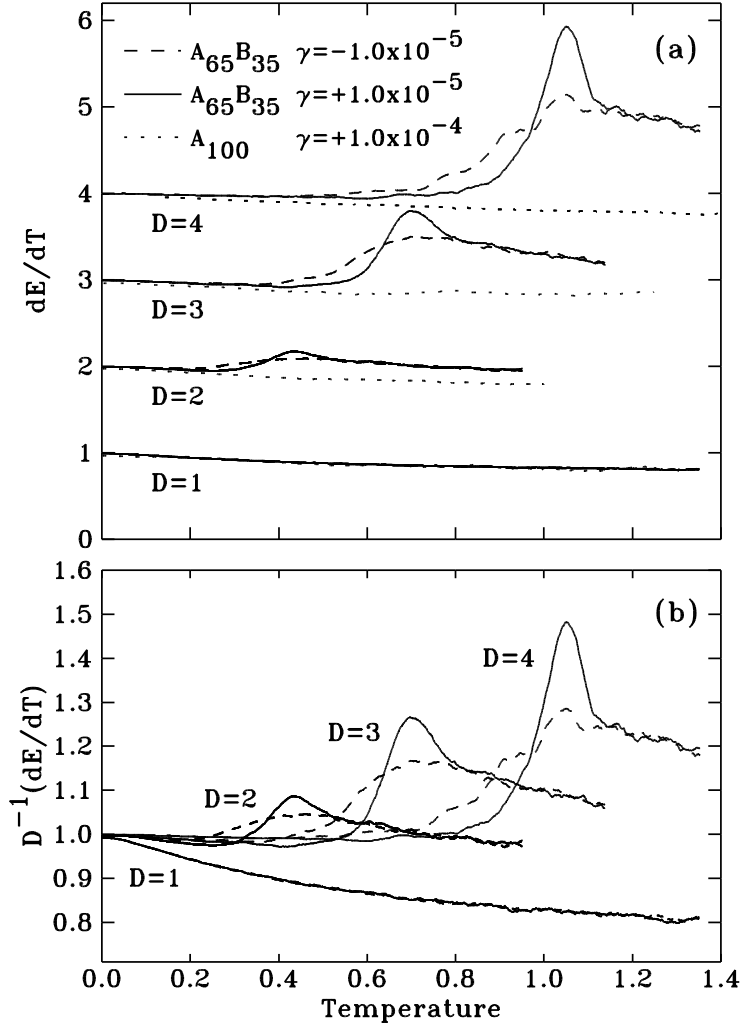}
\caption{\label{stack_1} 
(a) Constant volume specific heat, per particle, for A$_{65}$B$_{35}$
upon cooling (dashed lines) followed by  heating (solid lines) at
$\gamma=\mp1.0 \times 10^{-5}$.  The dotted lines show the specific heat
of crystallized A particles upon heating at $\gamma=+1.0 \times 10^{-4}$.
The cooling data for the crystals (not shown) coincide with the data upon
heating. (b) Same data as in (a) but now divided by the dimension $D$.
}
\end{figure}

The specific heat at constant volume, $c_V=dE/dT$, is shown in
Fig.~\ref{stack_1}a. The law of Dulong-Petit requires that for classical
solids $ \displaystyle{\lim_{T\to 0}}c_V(T)=D$~\cite{ashcroft76} and we
find that the specific heat becomes indeed equal to $D$ as temperature
approaches zero.  Also included in the graph is the $T-$dependence of the
specific heat of the A$_{100}$ crystals.
At low $T$ the specific heat of the glasses and and the
crystals decreases with increasing temperature.  This decrease is due
to the anharmonicity of the effective potentials near the equilibrium
positions of the particles. To see this we consider this effect in
detail for the $D=1$ case.  For a chain segment with only one type of
particle the first three terms of a Taylor expansion of the effective
interparticle potential are given by $U(r) = U_0 + sr^2/2 + \lambda r^4$,
where $U_0 < 0$ is the binding energy at $T=0$.  The quadratic term,
with $s > 0$, leads to the nearly harmonic motion of the particle around
its equilibrium position that gives rise to the Dulong-Petit result.
The leading anharmonic term has a positive coefficient ($\lambda > 0$) due
to the repulsive interaction of the Lennard-Jones potential as neighboring
particles approach.  For a one-particle anharmonic oscillator of mass $m$
the classical limit of the quantum-mechanical result~\cite{aly00} is

\begin{equation} 
\label{anharm} 
C = 1 - 6 \lambda (m/s)^2 k_B T.
\end{equation} 

\noindent
The negative term shows that the specific heat decreases in
the vicinity of $T=0$. However, the one-particle calculation on
which Eq.~(\ref{anharm}) is based overestimates the decrease of
the specific heat of the $D=1$ particle chain by about a factor of
two. This discrepancy is due to the multi-particle effects in a linear
chain. Although it is unfortunately not possible to take into account
these effects in an exact way, there exist approximation schemes to
calculate them~\cite{westera75,cowley83} and it is found that at low $T$
the specific heat does indeed decrease linearly with increasing $T$. These
calculations also show that such a $T-$dependence is only found in the
specific heat at constant volume, whereas the one at constant pressure
increases with increasing $T$~\cite{cowley83}.

Figure~\ref{stack_1}a shows that for $D=1$ the specific heat of the
disordered glass coincides within the numerical accuracy of the data
with the one from the crystal, i.e. the anharmonic effects in the
two systems are very similar, at least in the $T-$range considered.
The $c_V(T)$ data for the
($D=1$) glass agrees with the one of the corresponding crystal not only
in the $T-$range in which there is a linear $T-$dependence, but also at
temperatures at which $c_V(T)$ is no longer linear. This is thus evidence
that these two systems have also similar higher order anharmonic effects.

Similar results are obtained for the case of two dimensions for which
we can compare the specific heat of the glass at low temperatures
with that of a hexagonal crystals consisting solely of A particles,
Fig.~\ref{stack_1}a.  Type A particles by themselves crystallize readily,
typically with about 2 vacancies per 1000 A particles, and the specific
heat { curves} upon cooling and heating
at $\gamma = \mp 10^{-4}$ agree with
each other (data not shown). We see that up to $T=0.2$ the specific heats
of the hexagonal A phase and A$_{65}$B$_{35}$ glass nearly coincide.
Therefore we can conclude that the linear decrease of the specific
heat is not just a particularity of the glassy state, but instead a
general property of both types of condensed Lennard-Jones systems at low
temperatures. Qualitatively the same results are obtained for $D=3$ and
$D=4$. This observation is in agreement with experimental findings since
there it has been found that it is advantageous to crystallize samples
{\it in situ} after measuring their specific heats in the supercooled
liquid and glassy states. By subtracting the measured specific heat of
the crystal one obtains the net glass and supercooled liquid signals,
and the net glass signal is typically indistinguishable from zero
\cite{bruning01}.

The excess specific heat is defined as $\Delta c_V(T) =
c_V^\text{liquid}(T)-c_V^\text{glass}(T)$, where $c_V^\text{liquid}$
is the specific heat of the system on the (metastable) liquid branch
and $c_V^\text{glass}$ is the specific heat of the frozen (non-ergodic)
glass. At $T_g$ we observe a step $\Delta c_V(T_g)$ as the system
switches between the two states.  This step and the hysteresis between
heating and cooling characterizes the glass transition. As with the
$P(T)$ curves, see Fig.~\ref{pressure}, we find no step or hysteresis
in $\Delta c_V$ for $D=1$. However, for $D \ge 2$ the glass transition
can be readily be identified, and we see that $T_g$ increases as the
dimensionality increases.  This rise of $T_g$ with $D$ is consistent
with the higher binding energies per nearest neighbor particle pair
(Fig.~\ref{pstack_E}). Also the amplitude of the hysteresis increases with
$D$, Fig.~\ref{stack_1}a. In Fig.~\ref{stack_1}b we normalize $c_V(T)$
by $D$, the specific heat at $T=0$.  Even with this normalization both
$\Delta c_V(T_g)$ and the area of the hysteresis loop increase with
$D$,  reflecting the larger number of steric degrees of freedom when
the particles can move in more dimensions.

Figure~\ref{stack_1} also shows that above $T_g(D)$ the shape of the
$c_V(T)$ curves are independent of $D$.  This similarity of the specific
heat suggests that, in terms of their thermal fluctuations, $D=1$
particle chains behaves like a fluid down to $T=0$. This observation is
in agreement with analytical calculations for soft-sphere systems which
find that 2.0 is the minimum (fractional) dimension required for a glass
transition \cite{thalmann02}.   { The absence of a transition
for $D=1$ may be linked to the fact that the A$_{65}$B$_{35}$ chains
are non-erodic, since the initial random order of the particles in the chains
remains fixed.  Upon cooling these chains cannot reach an energetically
favorable state with a higher number of nearest A--B neighbors, while
the systems in higher dimensions can reach such chemically more ordered
states.}

\begin{figure}[top]
%location /puffin/ralf/everything/ljcnh/pstack3b
\includegraphics[scale=0.9]{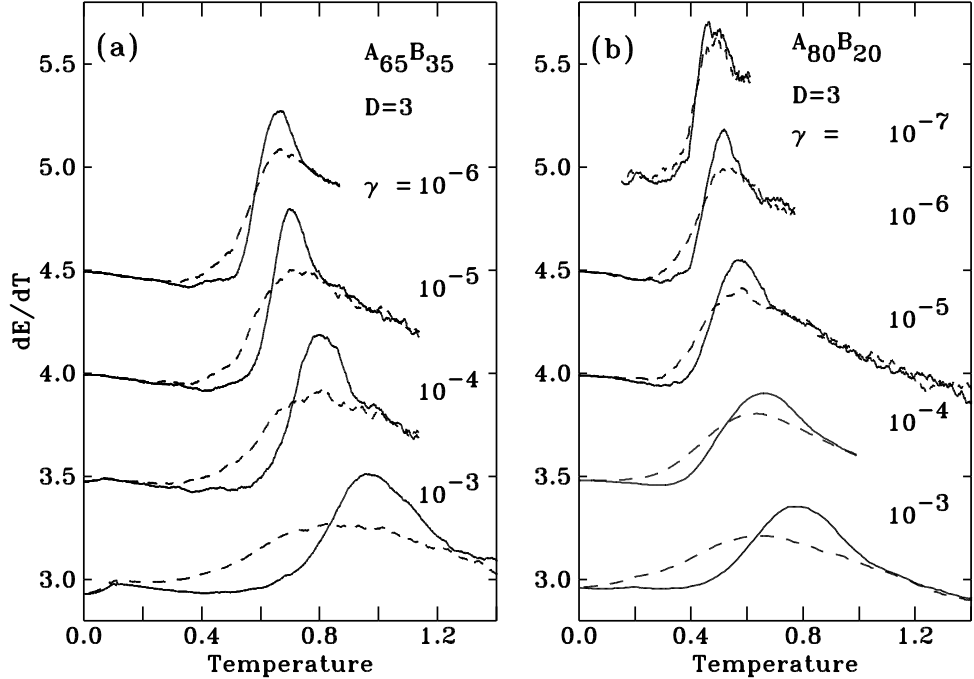}
\caption{\label{pstack3b}
Constant volume specific heat, per particle,  of (a) A$_{65}$B$_{35}$
and (b) A$_{80}$B$_{20}$ for $D=3$ upon cooling (dashed lines) followed
by heating (solid lines).  Curves at rates less than $\gamma = 10^{-3}$
are shown successively shifted up by 0.5. 
}
\end{figure}

\begin{figure}[top]
%location /goose/denis/ljcnh/pstack3c
\includegraphics[scale=0.9]{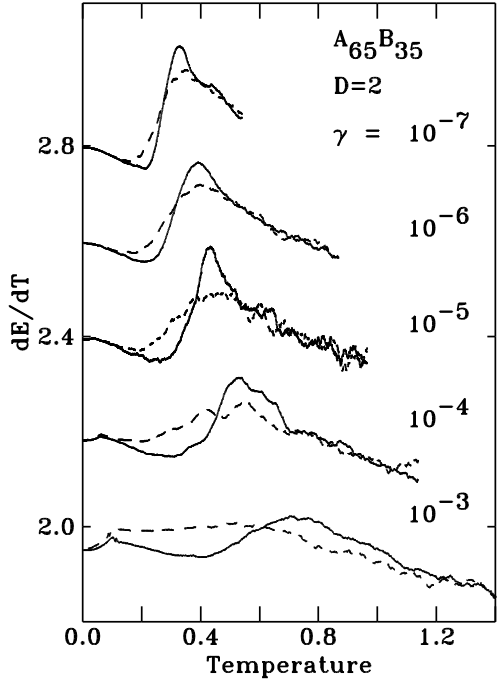}
\caption{\label{pstack3c}
Constant volume specific heat, per particle,  for A$_{65}$B$_{35}$ in two
dimensions upon cooling (dashed lines) followed by heating (solid lines).
Curves at rates slower than $\gamma = 10^{-3}$ are shifted up successively
by 0.2.
}
\end{figure}
	
\begin{figure}[top]   
%not at location goose/denis/ljcnh/p_expthe
%puffin /Users/ralf/everything/ljcnh/tests/tests/tftest
\includegraphics[scale=0.9]{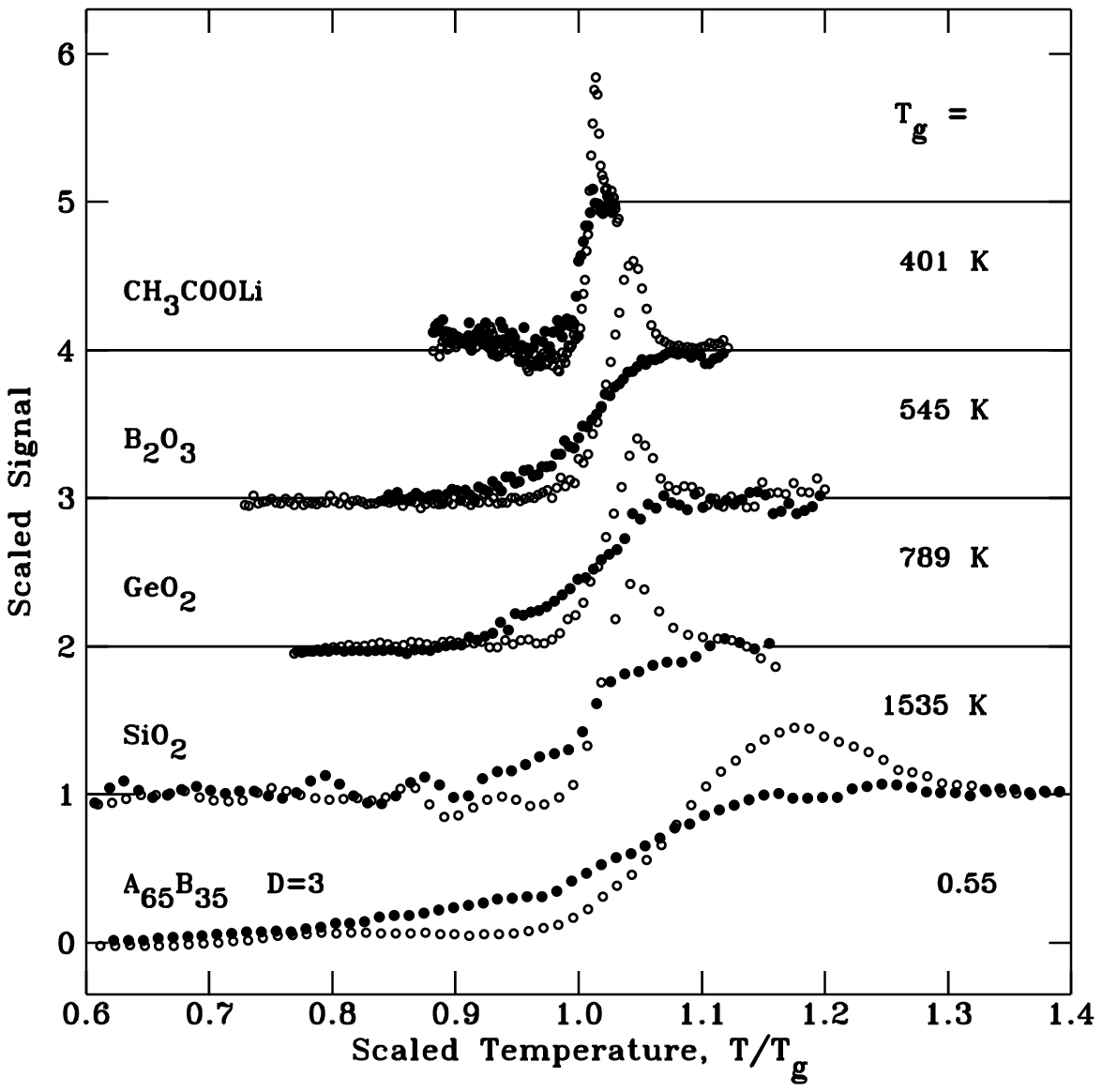}
\caption{\label{p_expthe}
Comparison of the $D=3$, A$_{65}$B$_{35}$, $\gamma=10^{-6}$
simulation result with the glass transition in  experimental
systems~\cite{bruning96,bruning03,bruning05}. For lithium acetate and
GeO$_2$ the data are scaled results from calorimetric experiments, while
for B$_2$O$_3$ and SiO$_2$ the curves are based on volume expansion and
small angle x-ray scattering data, respectively.  Temperature:s are scaled
by $T_g$. Ordinate values are scaled by  subtracting the (extrapolated)
signal of the glass state and dividing the residual by the (extrapolated)
supercooled liquid signal.  Curves other than the BMLJ system are shifted
up by successive steps of 1.
}
\end{figure}

Since the glass transition is related to the fact that the
system falls out of equilibrium, the specific heat curves will, at
temperatures around $T_g$, depend on the cooling and heating rate. In
Fig.~\ref{pstack3b}a we show the specific heat of the A$_{65}$B$_{35}$
mixture in three dimensions using temperature scanning rates that
are varied by three orders of magnitude. As in real experiments
\cite{ritland54,yang87,limbach88,johari89,bruning92,bruning94},
$T_g$ decreases with cooling rate. In Fig.~\ref{pstack3b}b we show
$c_V(T)$ for the A$_{80}$B$_{20}$ composition, using scanning rates
that vary by four decades. Although at a first glance the data
for the two compositions look quite similar, there are significant
differences. Firstly in the A$_{65}$B$_{35}$ system the difference
between the heating and cooling curves is larger than in the case of
A$_{80}$B$_{20}$.  This difference may be related, see the discussion
of Fig.~\ref{pGlass}, to a higher concentration of locally frustrated
structures in the A$_{65}$B$_{35}$ glass, and thus to its better glass
forming ability and resistance against crystallization. Secondly the
A$_{80}$B$_{20}$ data has, for the three slowest-rate cooling curves, a
unusual feature that is not present in $c_V(T)$ for the A$_{65}$B$_{35}$
system in that one sees a small peak in the specific heat that coincides
with the maximum of the heating curve, see Fig.~\ref{pstack3b}b. This
feature could be a sign of incipient crystallization, e.g.\ due to the
appearance of sub-critically sized nuclei.

Since $D=2$ simulations require less computational effort that the three
dimensional systems, slower cooling and heating rates can be investigated.
Figure~\ref{pstack3c} shows the $D=2$, A$_{65}$B$_{35}$ system
for scanning rates that vary by four decades and we see that all of them
have a well defined glass transition.  We note a substantial decrease
of the width of the glass transition, in agreement with the data shown
in Fig.~\ref{pstack3b}. In fact, in general there is no qualitative
difference between this system and the one in $D=3$ and hence we can
conclude that qualitatively the glass transition does not depend on $D$,
if $D\geq 2$.

From Figs.~\ref{pstack3b} and \ref{pstack3c} one recognizes
clearly that above $T_g$ the specific heat increases with decreasing
temperature, in agreement with previous studies of this Lennard-Jones
system~\cite{vollmayr96,sciortino00}. Such a $T-$dependence is not
unexpected since the relaxation time shows a significant non-Arrhenius
dependence on $T$~\cite{kob94,kob95a,kob95b}, i.e. the system can be
considered as ``fragile''~\cite{bohmer94}, and experimentally it is known
that fragile glass-formers usually have a specific heat that increases
with decreasing $T$~\cite{alba90,angell94}.

Due to this increase of $c_V$ with decreasing $T$, the excess specific
heat $\Delta c_V(T) = c_V^\text{liquid}(T)-c_V^\text{glass}(T)$
becomes larger at lower temperatures, in accordance with experimental
results~\cite{angell91}. Accordingly the step $\Delta c_V(T_g)$
becomes larger as $T_g$ decreases upon slower cooling. Furthermore
we see from Figs.~\ref{pstack3b} and \ref{pstack3c} that the heat
flow rises above the specific heat of the supercooled liquid as the
glass regains metastable equilibrium when the system is heated to $T >
T_g$, an effect that also observed in real experiments and which is
due to the kinetics of the glass transition.  We compare the
$T-$dependence of the specific heat found in the present simulations
with experimental results by using normalized scales.
Following Tool, we scale the ordinate by converting the specific heat signal
into the fictive temperature $T_f(T)$ of the system~\cite{tool46}. For
this we have linearly extrapolated the specific heat of the glass to the
supercooled liquid regime, and the specific heat of the supercooled
liquid was extrapolated into to glass regime~\cite{sciortino00}. One
then subtracts the (extrapolated) glass signal from the data as well as
from the supercooled liquid curve. Finally one divides the residual
curves (cooling and heating) by the residual supercooled liquid
curve.  The scaled data $dT_f/dT$ is shown as a function of $T/T_g$
in Fig.~\ref{p_expthe}.  Here $T_g$ was defined consistently as the
value of $T_f$ reached upon cooling to the lowest temperature.
Note that by construction the scaled signal
goes to zero at low temperatures and to unity at high temperatures.
Also included in the graph is the corresponding data from experiments
of different glass-formers~\cite{bruning96,bruning03,bruning05}. The
comparison of the data from simulation with the one from experimental
shows that the hysteresis at the glass transition of the BMLJ system
reproduces indeed the shape observed in experiments carried out with
fragile and strong glass-formers. We also see that the width of the
glass transition as observed in the simulation is significantly larger
than the one found even in the strongest glass-forming substances,
such as SiO$_2$, and much larger than the fragile glass-former lithium
acetate. This difference could have been expected since this width
depends not only on the fragility of the glass-former but also on the
cooling rate~\cite{vollmayr96}, and the temperature scanning rates of
the simulations are typically $10^{10}$ times faster than laboratory
rates~\cite{ritland54,bruning92}.

\begin{figure}[top]	           
% goose/denis/ljcnh/petot_zoom
\includegraphics[scale=0.7]{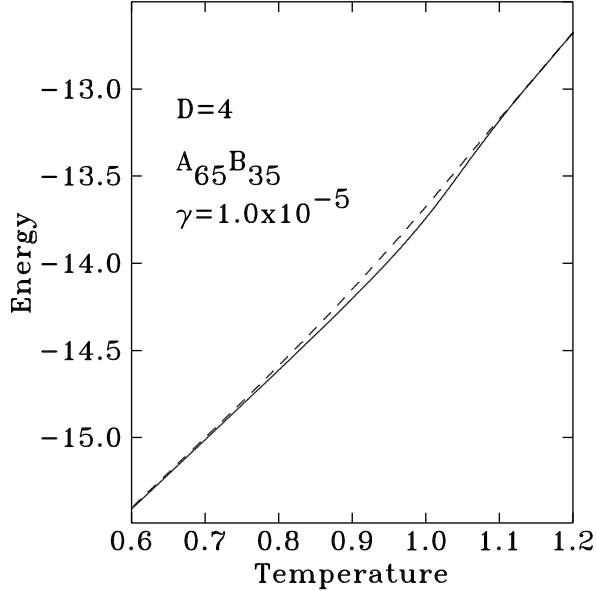}
\caption{\label{petot_zoom} 
The total energy per particle for A$_{65}$B$_{35}$ in four dimensions
upon cooling (dashed line) followed by heating (solid line) at rate
$\gamma=\mp 1.0\times 10^{-5}$.
}
\end{figure}

\begin{figure}[top]   
%location goose/denis/ljcnh/p_arehei
\includegraphics[scale=0.9]{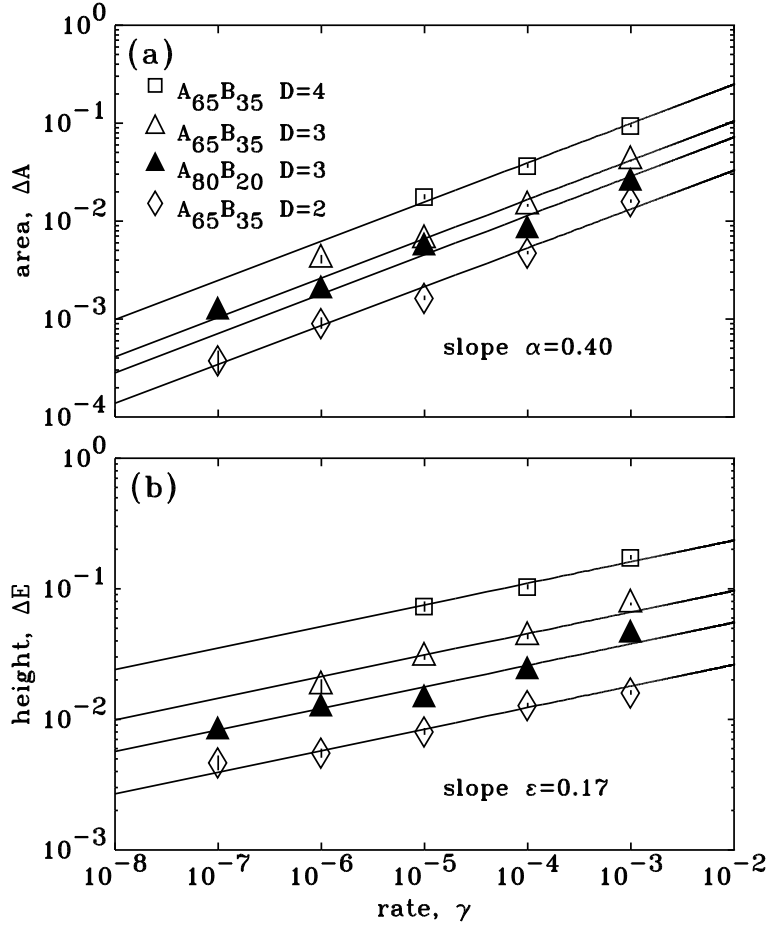}
\caption{\label{p_arehei}
Area, (a), and height, (b), of the hysteresis loop at the glass transition
versus the scanning rate on logarithmic scales for different compositions
and number of spatial dimensions.  Symbols correspond to the systems
presented in the upper left corner of (a).  Solid lines are fits to
the power laws $\Delta A= \Delta A_{0}(\gamma/10^{-5})^\alpha$ for (a)
and $\Delta E= \Delta E_0(\gamma/10^{-5})^{\epsilon}$ for (b).
}
\end{figure}

In the following we consider the dependence of the hysteresis on
temperature scanning rate, composition, and dimensionality of the system
in more detail.  Figure~\ref{petot_zoom} shows the total energy $E(T)$
upon cooling and heating for the four-dimensional A$_{65}$B$_{35}$
system.  We denote the area of the hysteresis loop between heating and
cooling curves as $\Delta A$, and their maximum separation as $\Delta
E$. Figure~\ref{p_arehei} shows these two quantities as a function
of temperature scanning rate $\gamma$ on logarithmic scales. The data
points fall, to a good approximation, onto parallel lines, indicating a
power-law dependence of $\Delta A$ and $\Delta E$ on $\gamma$. Therefore
we fitted the data with the functional form $\Delta A(\gamma)=\Delta
A_0(\gamma/10^{-5})^{\alpha}$ and  $\Delta E(\gamma)=\Delta
E_0(\gamma/10^{-5})^{\epsilon}$, where $\Delta A_0$ and $\Delta E_0$ are
the area and the maximum separation at $\gamma=10^{-5}$, respectively.
The exponents that best fit all data are $\alpha=0.40 \pm 0.02$ and
$\epsilon=0.17 \pm 0.03$. Table~\ref{tab1} lists the pre-factors $\Delta
A_0$ and $\Delta E_0$. From this table, and from Fig.~\ref{p_arehei},
we can conclude that the hysteresis effect increases with increasing $D$,
in agreement with the result shown in Fig.~\ref{stack_1}, i.e. that the
glass-forming ability of the system increases with its dimensionality.

A similar analysis can also be done for real systems, although of
course only for $D=3$~\cite{bruning03}. It is found that the area and
maximum separation of the hysteresis curves also show a power-law
dependence on the cooling rate, thus showing that the results from
the present simulations are consistent with real experiments.  However,
these experiments yield exponents of $0.29$ for area and $0.15$ for the
height, i.e. different values from the one found here and hence we can
conclude that these exponents are not universal, but material-specific
or characteristic of the regime of temperature scanning rate.

Since fragile glass-formers systems have relaxation times $\tau(T)$
that seem to diverge at a finite temperature \cite{bohmer94}, their
effective activation barriers, i.e. the local slope of $\log(\tau)$
vs. $1/T$ is larger than the one for strong glass-formers, if $\tau$
has macroscopic values, e.g. 1~second. The temperature range at which
the system falls out of equilibrium at the glass-transition is therefore
trivially related to the fragility of the system, with fragile (strong)
glass-formers showing a transition in a narrow (wide) temperature range.
As mentioned above, the present BMLJ system is expected to be a fragile
glass-former since its relaxation times show a strongly non-Arrhenius
$T-$dependence. Using the power-laws approximations for the scanning
rate dependence of the hysteresis loop (Fig.~\ref{p_arehei}), we
can obtain a rough estimate for the behavior of the BMLJ systems at
laboratory scanning rates ($\gamma = 10^{-15} \approx 2 {\rm K/s}$,
if we identify the A particles as argon atoms).
Of course it is highly
uncertain whether or not the present BMLJ system avoids crystallization
at such slow rates.  The width of the glass transition $\Delta T$ is
proportional to the ratio $\Delta A / \Delta E$ and thus its
$\gamma$-dependence is given by the exponent $\alpha - \epsilon$.  Referring to
Fig.~\ref{p_expthe}, one can estimate the width of the hysteresis loop
for A$_{65}$B$_{35}$ at $\gamma = 10^{-6}$ to be $\Delta T = 0.5 T_g$.
The power-law approximation, with $\alpha - \epsilon \approx 0.23$,
predicts then for the laboratory rate a $\Delta T = 0.004 T_g$ (or $\Delta
T = 0.027 T_g$ if using the experimentally derived values $\alpha =
0.29$ and $\epsilon = 0.15$). By comparison, the width of the (fragile)
lithium acetate glass curve in Fig.~\ref{p_expthe} is about 0.04.
We conclude that, based on this analysis, the BMLJ system is indeed
fragile, in agreement with the data from the $T-$dependence of the
relaxation times~\cite{kob94,kob95a,kob95b}. Last not least we also can
conclude that the fragility depends on the dimensionality of the system,
since the width of the glass transition increases with increasing $D$
(see Fig.~\ref{p_arehei}).

\begin{table}
\begin{center}	 
\begin{tabular}{l  r @{$\pm$}l r@{$\pm$}l}
\hline\hline
% \cline{2-5}
\multicolumn{1}{c}{System} &\multicolumn{2}{c}{$\Delta E_0 \times 10^{3}$}&\multicolumn{2}{c}{$\Delta A_0 \times 10^3$}\\
\hline
A$_{65}$B$_{35}$, $D=2$\hspace{5mm}\ &$2.1$&$0.4 $\hspace{5mm}\ &$8.4$&$0.7$\\
A$_{65}$B$_{35}$, $D=3$&$6.6$&$0.8$&$31 $&$5$\\
A$_{65}$B$_{35}$, $D=4$&$16.0$&$1.5$&$76 $&$6$\\
A$_{80}$B$_{20}$, $D=3$&$4.5$&$0.2$&$18 $&$2$\\
\hline\hline
\end{tabular}
\end{center}
\caption{Values of the pre-factor for the power-law fits $\Delta
E = \Delta E_0(\gamma/10^{-5})^{0.17}$ and $\Delta A = \Delta
A_0(\gamma/10^{-5})^{0.40}$, representing the height and area of the
hysteresis loop in the total energy, respectively.}
\label{tab1}
\end{table}
	
\section{Summary}	
We have presented the results of molecular dynamics simulations in order
to study how the calorimetric glass transition of binary Lennard-Jones
systems depends on the cooling rate and the number of dimensions. The
BMLJ systems were cooled at a constant rate $-\gamma$, followed by
reheating to the ergodic liquid state at rate $\gamma$. We find that
the composition A$_{80}$B$_{20}$, a good glass-former in
three dimensions, crystallizes in two dimensions if $\gamma$ is small,
whereas the  composition A$_{65}$B$_{35}$ does not crystallize
for any dimension at the cooling rates investigated here.

For glasses that have been produced with a given (small) $\gamma$ we find
that the peaks in the radial distribution functions become quickly washed out
with increasing dimensionality $D$, reflecting fewer geometric
constraints.  In particular the nearest neighbor peak of the
radial distribution functions becomes broader, reflecting a wider range
of geometric configurations in higher dimensions.
This is evidence that the glass-forming ability of the system
increases with increasing dimensionality.  As $D$ increases, the first
peak in the radial distribution function shifts closer to the central
atom due to force exerted by the second-nearest neighbors on the nearest
neighbors, thus resulting in a stronger $T-$dependence of the pressure. At
low temperatures the $T-$dependence of the constant volume specific heat
of the glass is very close to the one of a one-component Lennard-Jones
crystal, showing that the anharmonic effects in glasses and crystals
are quite similar.

A glass transition was observed in two, three and four dimensions,
whereas no glass transition is observed in one dimension. For $D=1$
the specific heat curve resembles at all temperatures the one of
the supercooled liquid for $D\ge2$, and thus we conclude that the one
dimensional system behaves kinetically like a liquid down to $T=0$.

For the systems that show a glass-transition we find a hysteresis loop
(cooling and heating cycle) in the energy per particle as well as in
the pressure of the system.  At a given cooling rate the area of this
loop and the temperature at which it occurs increase with increasing
$D$. Thus this is further evidence that increasing dimensionality raises
the glass-forming ability of the system.

The glass transition becomes sharper with decreasing temperature scanning
rate and thus the hysteresis loop shrinks.  Power-laws can be used to
fit scanning rate dependence of the area and height of these loops.
The exponents describing this $\gamma$-dependence seem to be independent
of composition or dimensionality. A similar analysis of experimental
data indicate that these exponents are not universal, but appear to
be specific for the system considered.

\section{Acknowledgments}	

The Natural Sciences and Engineering Research Council of Canada
(NSERC) has supported this work through a discovery grant. S.P. and
D.St-O. especially thank NSERC for Summer Undergraduate Research Awards.
We gratefully acknowledge the Mount Allison Cluster for Advanced Research
(TORCH) for allocations of computer resources.

\end{document}